\documentclass[11pt, a4paper]{article}
\usepackage[margin=1in]{geometry}
\usepackage{amsmath,amssymb}

\usepackage{changepage}

\usepackage{textcomp,marvosym}
\usepackage[square,numbers]{natbib}
\usepackage{bm}
\usepackage{bbm}
\usepackage{graphicx}
\usepackage{enumerate}
\usepackage{amsmath}
\usepackage{comment}
\usepackage{algorithm}
\usepackage{algpseudocode}
\usepackage{multirow}
\usepackage{cite}

\usepackage{nameref,hyperref}

\usepackage{subcaption}
\usepackage[right]{lineno}

\usepackage[nopatch=eqnum]{microtype}
\DisableLigatures[f]{encoding = *, family = * }

\usepackage[table]{xcolor}

\usepackage{array}
\global\long\def\mbeta{\boldsymbol{\beta}}%
\global\long\def\mgamma{\boldsymbol{\gamma}}%
\newcolumntype{+}{!{\vrule width 2pt}}

\makeatletter
\renewcommand{\@biblabel}[1]{\quad#1.}
\makeatother

\begin{document}
	\vspace*{0.2in}
	
	\begin{flushleft}
		{\Large
			\textbf{Integrating complementary datasets for constructing microbiome-metabolome-disease pathways} 
		}
		\newline
		\\
		\centering
		Lei Fang\textsuperscript{1},
		Chenglong Ye\textsuperscript{2},
		Yue Wang\textsuperscript{3*},
		\\
		\bigskip
		\textbf{1} Department of Statistics, Miami University, Oxford, Ohio, USA
		\\
		\textbf{2} Dr. Bing Zhang Department of Statistics, University of Kentucky, Lexington, Kentucky, USA
		\\
		\textbf{3} Department of Biostatistics and Informatics, University of Colorado Anschutz Medical Campus, Aurora, Colorado, USA
		\\
		\bigskip
		
		*Correspondence author: yue.2.wang@cuanschutz.edu
		
	\end{flushleft}
	
	
	%
	\section*{Abstract}
    Microbiome–metabolome–disease pathways play a central role in host health, yet their study is hindered by limited data availability: large microbiome cohorts often lack metabolomic measurements, while multi-omics studies with paired profiling are typically small or incompletely phenotyped. 
    We propose an integrative statistical framework that integrates complementary datasets, a large clinical cohort with microbiome data but missing metabolomics, and a smaller external study with detailed metabolomic profiling, to enable pathway-level inference.
    Our approach is based on a high-dimensional structural equation model that jointly characterizes {relationships} among the microbiome, metabolome, and disease outcome, while grouping microbes into functional categories. We address the compositional structure of microbiome data via the additive log-ratio (ALR) transformation and perform estimation using penalized two-stage least squares. 
    A key innovation is a novel multiple-reference inference procedure that enables valid testing of metabolome--disease associations and is robust to the choice of reference in the ALR transformation. 
    Simulation studies demonstrate favorable finite-sample performance. An application to inflammatory bowel disease data reveals biologically interpretable microbiome–metabolome–disease links.
	
\noindent \textbf{Keywords} {high-dimensional regression, robust inference, stability selection, two-stage least
squares}

	
	%

	\section*{Introduction}
	\label{sec: intro}
	
	\sloppy 
	Since the initiation of the human microbiome project in 2007 \citep{turnbaugh2007human}, mounting evidence has suggested that the human microbiome plays a crucial role in maintaining our health and well-being by aiding in digestion, immune system regulation, and synthesizing essential nutrients \citep{brune2014symbiotic,benson2010individuality,clarke2014minireview}. 
	It has also been associated with various diseases and conditions, including inflammatory bowel disease (IBD), allergic disorders, Type 1 diabetes, autism, obesity, and colorectal cancer \citep{lloyd2019multi,johansson2011early,alkanani2015alterations,de2013fecal,ley2005obesity,dejea2014microbiota}. 
	However, the underlying mechanism by which the gut microbiome interacts with host health is still poorly understood.
	
	Microbial metabolomics in the human gut opens a new avenue to investigate the host-microbiome {relationship}. 
	Recent studies identified various microbiome-metabolome {relationships} under different conditions. For example, gut microbiota-derived metabolites play central regulator roles in metabolic disorders \citep{lamichhane2018gut,agus2021gut}.
	Studies have also identified gut metabolites significantly associated with host health \citep{visconti2019interplay,rahman2023gut}. 
	These results jointly suggest critical microbiome-metabolome-host pathways for maintaining host health. 
	
	However, several data limitations hinder the comprehensive investigation of microbiome-metabolome-host pathways. Ideally, a single dataset would have a sufficiently large sample size, comprehensive microbiome and metabolome profiling, and relevant clinical outcomes to enable robust pathway analysis. 
	However, in reality, such datasets are rare. Most existing microbiome studies lack paired metabolomics data, preventing the joint analysis of microbiome, metabolome, and phenotype relationships. While emerging multi-omics technologies \citep{quince2017shotgun} allow simultaneous profiling of gut microbiome and metabolome from the same stool sample, these datasets remain limited in number and often suffer from small sample sizes or a narrow range of clinical outcomes, reducing statistical power and biological interpretability.
	Further complicating this issue is the diversity of metabolomics platforms. Differences between targeted vs. untargeted approaches, analytical techniques, and metabolite coverage lead to substantial variability in the types and numbers of metabolites measured across studies \citep{bingol2018recent}. As a result, key metabolites of interest may not be captured in existing paired microbiome-metabolome datasets, making pathway reconstruction challenging.  
	
	A potential approach to overcoming these challenges is to conduct integrative analysis, where complementary datasets are combined to construct microbiome-metabolome-host pathways. 
	Specifically, one dataset has a large sample size and relevant clinical outcomes but lacks metabolite measurements (called the target data), while another includes detailed metabolomic profiling but has a smaller sample size or lacks disease phenotype data (called the external data). 
    { 
    This setting presents several methodological challenges, including handling missing metabolite measurements across datasets, accommodating the high dimensionality of microbiome features, and accounting for the compositional nature of microbiome data.}
	 We develop a high-dimensional structural equation model (SEM) with compositional covariates to address these challenges that leverages complementary datasets. 
    Our SEM allows for the presence of hidden confounders that simultaneously influence gut metabolism and host health. 
    This ensures a more robust identification of the pathway, particularly given that many confounders in microbiome studies, such as dietary factors, are often unmeasured.

	The proposed SEM identifies {relationships} among the microbiome, metabolome, and disease outcome, while grouping microbes into functional categories, each representing a distinct function that the microbes play in the pathway. 
We address the compositional nature of the microbiome data via the additive log-ratio (ALR) transformation and develop a { high-dimensional} penalized two-stage least squares framework with predicted covariates to perform parameter estimation.  
    Specifically, the first stage employs an $\ell$-1 penalized regression to train a microbiome-metabolome {relationship} model using the external data, which is then used to predict the missing metabolite in the target data.  
	The predicted metabolite is further taken into the second stage to analyze its association with the phenotype while adjusting for microbial confounders. 
    We assess the statistical significance of metabolite–host associations using a regularized permutation test, 
    { which provides valid finite-sample inference in the high-dimensional setting}. 
    {Such microbiome-based ``predictive metabolic analysis'' across independent datasets has been shown to be successful in recent studies, such as MelonnPan \citep{mallick2019predictive}. 
    }
	
Although the ALR transformation effectively addresses the compositional nature of microbiome covariates, it has a well-known limitation: the results depend on the choice of a pre-selected reference microbe, and different references may lead to different conclusions. 
To address this reference-dependence issue, we further develop a robust multiple-reference procedure that treats each microbe in turn as the reference and integrates the resulting analyses. Specifically, we combine permutation-based p-values using a Cauchy combination test \citep{liu2020cauchy}, and apply stability selection \citep{meinshausen2010stability} to identify microbes within each functional category.

	
    
	
	Throughout the article, we use normal typeface to denote scalars, bold lowercase typeface to denote vectors, and uppercase typeface to denote matrices. For any vector ${\bf v} \in \mathbb{R}^p$, we use $v_l$ to denote the $l$-th entry of ${\bf v}$ for $l = 1, \ldots, p$, and for any set $\mathcal{I} \in \{1, \ldots, p\}$, ${\bf v}_{\mathcal{I}}$ denotes the sub-vector of ${\bf v}$ indexed by $\mathcal{I}$.
	The $l_q$ norm of ${\bf v}$ is defined as $\|{\bf v}\|_q=(\sum_{l=1}^p |v_l|^q)^\frac{1}{q}$ for $q > 0$ with $\|{\bf v}\|_0=|{1\le l\le p : v_l\ne 0}|$. 
	Let $\mathbbm{1}(\mathcal{A})$ denote the indicator function of the event $\mathcal{A}$; that is, $\mathbbm{1}(\mathcal{A}) = 1$ if $\mathcal{A}$ happens and $\mathbbm{1}(\mathcal{A}) = 0$ otherwise.

	\section*{Materials and methods}\label{sec: method}
	\subsection*{The target model}\label{sec: model} 
	We first define the true model in the target study. 
	Suppose that we observe the clinical outcome $y_i$ and the microbiome counts ${\bf w}_{i}=(w_{i1}, \ldots, w_{ip})^\intercal$ of $p$ taxa for $i=1,\dots,n$. 
	Letting $m_i$ denote the {\it unmeasured} gut metabolite of interest, we propose the following linear structural equation model (SEM): 
	\begin{align}
		&m_i= (\log({\bf w}_i))^\intercal\mgamma^0 +\delta_i, \; {\bf 1}_p^\intercal\mgamma^0=0  \label{sub model}\\
		&y_i=m_i\theta^* + (\log({\bf w}_i))^\intercal  {\mbeta}^0  + \varepsilon_i, \; {\bf 1}_p^\intercal\mbeta^0=0.  \label{whole model}
	\end{align}
	The constraints ${\bf 1}_p^\intercal\mgamma^0=0$ and ${\bf 1}_p^\intercal\mbeta^0=0$ are due to the compositional nature of microbiome measurement, following the general idea of the log-contrast model \citep{aitchison1982statistical,shi2016regression}. 
	To account for potential unmeasured confounding, we allow the error terms  $\delta_i$ and $\varepsilon_i$ to be correlated zero-mean random variables and assume they are independent of ${\bf w}_i$. 
	
	Model \eqref{sub model} and \eqref{whole model} jointly define the entire microbiome-metabolome-host pathway. 
	First, $\theta^*$ represents the direct effect of the metabolite on the outcome.
	Letting $\mathcal{A}=\{1,...,p\}$ index all microbes, $\mbeta^0 $ and $\mgamma^0$ jointly categorize the microbes into the following four groups: 
	
	\begin{itemize}
		\item $\mathcal{G}_1=\{l \in \mathcal{A} \:| \: \beta^0_l \ne 0 \: \textnormal{~and~} \: \gamma^0_l \ne 0 \}$: microbes in $\mathcal{G}_1$ are associated with both the outcome and metabolites;
		
		\item $\mathcal{G}_2=\{l \in \mathcal{A} \:| \: \beta^0_l = 0 \: \textnormal{~and~} \: \gamma^0_l \ne 0 \}$: microbes in $\mathcal{G}_2$ are associated with the outcome only through the metabolite;
		
		\item $\mathcal{G}_3=\{l \in \mathcal{A} \:| \: \beta^0_l \ne 0 \: \textnormal{~and~} \: \gamma^0_l = 0 \}$: microbes in $\mathcal{G}_3$ are directly associated with the outcome without affecting the metabolite; 
		
		\item $\mathcal{G}_4=\{l \in \mathcal{A} \:| \: \beta^0_l = 0 \: \textnormal{~and~} \: \gamma^0_l = 0 \}$: microbes in $\mathcal{G}_4$ are irrelevant to the host-metabolome {relationship}.
	\end{itemize}
	
	{Our goal is to make inference of $\theta^*$ as well as identify $\mathcal{G}_1, \mathcal{G}_2, \mathcal{G}_3, \mathcal{G}_4$.}
	However, models \eqref{sub model} and \eqref{whole model} cannot be directly fitted using the target data, where the metabolite \( m_i \) is not measured.  
	Below, we outline our approach for estimation and inference by leveraging an informative external dataset that includes measurements of the metabolite.

	\subsection*{Estimation and Inference}
	Our procedure proceeds in three steps.
	First, we handle the linear constraints in \eqref{sub model} and \eqref{whole model} via an additive log-ratio (ALR) transformation, using a pre-selected reference microbe \citep{aitchison1982statistical}.
	Second, we obtain estimates of $\bm{\beta}^{0}$, $\bm{\gamma}^{0}$, and $\theta^*$ through a penalized two-stage least squares approach.
	Third, we compute a p-value for testing $H_0:\theta^* = 0$ using a regularized permutation procedure.

	\subsubsection*{Step 1: The ALR transformation}
	

	The ALR transformation is based on a pre-selected reference microbe, e.g., $w_{ip}$. For each of the rest microbes, its ALR transformed abundance is defined as $x_{ij} = \log( w_{ij}/ w_{ip} )$ for $i = 1, \ldots, n$ and $j = 1, \ldots, p-1$. 
	Therefore, we can rewrite \eqref{sub model} and \eqref{whole model} with the ALR-transformed data as follows: 
	\begin{align}
		&m_i= {\bf x}_i^\intercal\mgamma^* +\delta_i,  \label{ALR sub model}\\
		&y_i=m_i\theta^* +  {\bf x}_i^\intercal  {\mbeta}^*  + \varepsilon_i, \label{ALR whole model}
	\end{align}
	where  ${\bf x}_{i}=(x_{i1}, \ldots, x_{i(p-1)})^\intercal$, and $\mgamma^*$ and ${\mbeta}^* \in \mathbb{R}^{p-1}$ are the subvectors of $\mgamma^0$ and $\mbeta^0$ excluding the $p$-th entry, respectively.
	Theoretically, using different reference microbes leads to the same log-constrast models in \eqref{sub model} and \eqref{whole model}. 
	To see this, we can write \eqref{sub model} as 
	\begin{align*}
		m_i
		&= \sum_{j\neq l} \log(w_{ij}/w_{il}) \gamma_j^* +\delta_i
		= \sum_{j \neq l}\log(w_{ij}) \gamma_j^*  - (\sum_{j \neq l}\gamma_j^*) \log (w_{il}) +\delta_i
	\end{align*}
	for any reference microbe $l = 1, \ldots, p$. 
	{This also holds for model \eqref{whole model}}. 

	
	\subsubsection*{Step 2: The penalized two-stage least squares} \label{sec: inference}
	Suppose we have an external dataset where we observe microbiome count data \(\widetilde{\mathbf{w}}_{j} = (\widetilde{w}_{j1}, \ldots, \widetilde{w}_{jp})^\intercal\) from the same \(p\) taxa as in the target data, along with the abundance of the metabolite of interest, denoted by \(\widetilde{m}_{j} \in \mathbb{R}\), for \(j = 1, \ldots, N\). 
	We also define the ALR-transformed data in the external study $\widetilde{x}_{jl} = \log(\widetilde{w}_{jl}/\widetilde{w}_{jp})$ for $j = 1, \ldots, N$ and $l = 1, \ldots, p-1$, and denote $\widetilde{\bf x}_j = (\widetilde{x}_{j1}, \ldots, \widetilde{x}_{j,p-1})^\intercal$. 
	We develop a two-stage least-squares procedure for parameter estimation. 
	In Stage 1, we aim to predict the unknown $m_i$ in the target data by training model \eqref{ALR sub model} with the external data. 
	Specifically, we estimate $\bm{\gamma}^*$ in \eqref{ALR sub model} with:
	\begin{align} \label{lasso: 1st stage}
		\widehat{\mgamma}= 
        \underset{\mgamma}
        {\text{argmin}}\biggl\{ \sum_{j=1}^N\dfrac{ (\widetilde{m}_j-  \widetilde{\bf x}_{j}^\intercal \mgamma)^2}{2N}+\lambda_\gamma \|\mgamma\|_1 \biggr\},
	\end{align} 
	where $\lambda_{\gamma} > 0$ is a tuning parameter. 
	In Stage 2, we replace the unobserved metabolite value \(m_i\) in model \eqref{ALR whole model} with its predicted counterpart \(\widehat{m}_i = \mathbf{x}_i^\intercal \widehat{\bm{\gamma}}\), for \(i = 1, \ldots, n\).
	Similarly, we obtain $\widehat{\bm{\beta}}$ by minimizing  
	\begin{align} \label{lasso: 2nd stage}
       \sum_{i=1}^n \frac{\left( y_i - \widehat{m}_i\theta- {\bf x}_i^\intercal \mbeta \right)^2}{2n}+\lambda_\beta \left\{\|\mbeta\|_1 + |\theta| \right\},
	\end{align} 
	where $\lambda_{\beta} > 0$ is a tuning parameter.

	
\subsubsection*{Step 3: Regularized permutation test}

    Having obtained $\widehat{\bm{\beta}}$ and $\widehat{\bm{\gamma}}$, we next test $H_0: \theta^* = 0$ to identify significant metabolome-disease associations, while accounting the microbes' effects.  
	Specifically, let 
	\(
	\widehat{S}_{\bm{\beta}} = \{j : \widehat{\beta}_j \neq 0\},
	\)
	where $\widehat{\beta}_j$ is the $j$-th component of the Lasso estimator $\widehat{\bm{\beta}}$ obtained from \eqref{lasso: 2nd stage}.  
	Based on this selected set, we form the submatrix ${\bf X}_{\widehat{S}_{\bm{\beta}}}$ of ${\bf X} = ({\bf x}_1,\ldots,{\bf x}_n)^\intercal$, obtained by retaining only the columns indexed by $\widehat{S}_{\bm{\beta}}$.
	With a tuning parameter $\lambda > 0$, we then construct the regularized hat matrix
	\[
	\widetilde{\bf H}_\lambda
	= 
	{\bf X}_{\widehat{S}_{\bm{\beta}}}
	\left(
	{\bf X}_{\widehat{S}_{\bm{\beta}}}^\intercal
	{\bf X}_{\widehat{S}_{\bm{\beta}}}
	+
	\lambda {\bf I}
	\right)^{-1}
	{\bf X}_{\widehat{S}_{\bm{\beta}}}^\intercal,
	\]
	and define the corresponding regularized orthogonal projection matrix
	\(
	\widetilde{\bf R}_\lambda = {\bf I} - \widetilde{\bf H}_\lambda.
	\)

	Let $\widehat{T}$ denote the Pearson correlation between the predicted vector $\widehat{\bf m}$ and the residual $\widetilde{\bf R}_\lambda {\bf y}$, where ${\bf y} = (y_1, \ldots, y_n)^\intercal$. 
	Next, for each $b = 1, \ldots, B$, we generate a permuted residual
	\(
	\widetilde{\bf R}_\lambda \left( {\bf P}_b \widetilde{\bf R}_\lambda + \widetilde{\bf H}_\lambda \right) {\bf y},
	\)
	where ${\bf P}_b$ is a random permutation matrix that is generated by randomly permuting the order of rows (or columns) of the identity matrix. 
	We then compute its correlation with $\widehat{\bf m}$, denoted by $T_b$.
	Finally, the p-value is defined as 
	\begin{align}\label{perm:p}
		\textnormal{p-value}=2(B+1)^{-1} \left(\text{min}\big\{\sum^B_{b=1} \mathbbm{1}(T_b\ge \widehat{T}),\sum^B_{b=1} \mathbbm{1}(T_b\le \widehat{T})\big\} + 1\right).
	\end{align}
    The above testing procedure is performed using the {\tt R} package {\tt phd} \citep{hemerik2021permutation}. 

	
	\subsection*{A Robust Multi-Reference Procedure}
	{
     While the previous three-step procedure depends on a chosen reference microbe, different reference choices may yield substantially different inference results. Moreover, there is generally no principled way to determine an optimal reference a prior. To avoid relying on an arbitrary reference, we develop a robust procedure that integrates results obtained by treating every microbe in turn as the reference.
    }
    This approach synthesizes the correlation estimates from all references by computing the median $\widehat{T}$, combining permutation p-values via a Cauchy combination test, and 
	applying stability selection to identify robust non-zero effects in $\bm{\beta}^{(0)}$ and $\bm{\gamma}^{(0)}$. The full algorithm is presented next.
	
	\begin{algorithm}
		\caption{Robust Multi-Reference Procedure}
		\label{alg:robust_procedure}
		\begin{algorithmic}[1]
			\Require Target data \(\{(y_i, \mathbf{w}_i)\}_{i=1}^n\), external data \(\{(\widetilde{m}_i, \widetilde{\mathbf{w}}_i)\}_{i=1}^N\), stability threshold $\tau$ 
			
			\For{\(j = 1\) \textbf{to} \(p\)}
			\State Use microbe \(j\) as the reference
			\State Obtain \(\widehat{\bm{\gamma}}^{(j)}\) via \eqref{lasso: 1st stage}
			\State Obtain \(\widehat{\bm{\beta}}^{(j)}\) via \eqref{lasso: 2nd stage}
			\State Compute correlation $\widehat{T}^{(j)}$ and the
            permutation test \(p\)-value \(p_j\) via \eqref{perm:p}
			\EndFor
			
			\State \textbf{Final correlation estimate:} \(\widetilde{T} \gets \text{median}\{\widehat{T}^{(1)}, \dots, \widehat{T}^{(p)}\}\)
			
			\State \textbf{Final significance test:} Combine \(\{p_j\}_{j=1}^p\) via Cauchy combination method to obtain final \(p\)-value \(\widetilde{p}\) for \(H_0: \theta^* = 0\)
			
			\State \textbf{Stability selection for \(\bm{\beta}^{0}\) and \(\bm{\gamma}^{0}\):}
			\For{each microbe \(k = 1, \dots, p\)}  \Comment{For each \(k\), sum over all references \(j \ne k\)}
			\State Define selection frequency \(\pi_k^{(\gamma)} \gets \frac{1}{p-1} \sum_{\substack{j \ne k}} \mathbbm{1}\big(\widehat{\gamma}_k^{(j)} \ne 0\big)\)
			\State Define selection frequency \(\pi_k^{(\beta)} \gets \frac{1}{p-1} \sum_{\substack{j \ne k}} \mathbbm{1}\big(\widehat{\beta}_k^{(j)} \ne 0\big)\)
			\EndFor
			
			\State \textbf{Robust group categorization:}
			\For{each microbe \(k = 1, \dots, p\)}
			\If{\(\pi_k^{(\beta)} > \tau\) and \(\pi_k^{(\gamma)} > \tau\)}
			\State Assign \(k\) to \(\mathcal{G}_1\)
			\ElsIf{\(\pi_k^{(\beta)} \le \tau\) and \(\pi_k^{(\gamma)} > \tau\)}
			\State Assign \(k\) to \(\mathcal{G}_2\)
			\ElsIf{\(\pi_k^{(\beta)} > \tau\) and \(\pi_k^{(\gamma)} \le \tau\)}
			\State Assign \(k\) to \(\mathcal{G}_3\)
			\Else
			\State Assign \(k\) to \(\mathcal{G}_4\)
			\EndIf
			\Comment{\(\tau \in (0.5, 1)\) is a stability threshold}
			\EndFor
			
			\State \Return \(\widetilde{T}\), \(\widetilde{p}\), \(\{\mathcal{G}_1,\mathcal{G}_2,\mathcal{G}_3,\mathcal{G}_4\}\)
		\end{algorithmic}
	\end{algorithm}

	\newpage 
	\section*{Results}
	
	\subsection*{A data-driven simulation study} \label{numerical study}
	We used the metagenomics data from \citep{yachida2019metagenomic} as the target data with $n=347$ (referred to as ``YACHI") and the paired microbiome-metabolome data from \citep{franzosa2019gut} as the external data with $N=220$ (referred to as ``FRAN"). 
	For the target dataset, the relative abundance of 11,318 microbes at the genus level were original measured. Due to the excessive zeros (85\%), we first aggregated the microbial genera into 1,140 microbial orders and then removed microbial orders with more than 30\% of zeros across participants. There were 114 microbial orders remaining.  
	The external dataset contains the relative abundance data of 9,974 and 1,054 microbes at the genus level and order level, respectively. Among them, 112 microbial orders overlapped with the target dataset.
    To account for the remaining zeros, we added $10^{-8}$ to the relative abundance of the overlapping 112 microbes in both target and external datasets for the ALR transformation.

	We generated the metabolite abundance $m_i$ in the target data according to model \eqref{sub model} based on the 112 overlapping microbial orders,
	where 
	$\gamma^0_j=0.5$ for $j=1, 2$  and 4, $\gamma^0_j=-0.75$ for $j=3$ and 5, and $\gamma^0_j=0$ for $j>5$.
	We then simulated the outcome $y_i$ according to model \eqref{whole model}, where $\beta^0_j =0.5$ for $j=4,6$ and 8, $\beta^0_j=-0.75$ for $j=5$ and 7, and $\beta^0_j=0$ for the rest.
	We set $\theta^*$ to various values to examine the effectiveness of the proposed method in detecting various levels of metabolite-disease associations. 
	Each pair $({\delta}_i, \varepsilon_i)$ were drawn from a bivariate normal distribution with mean zero, standard deviation $\sigma_\delta=4$ and $\sigma_\varepsilon=4$, and covariance 0.1, thereby allowing for unobserved confounding. 
    {The $R^2$ is approximately $23\%$ in model \eqref{sub model}, and the $R^2$ is approximately $9\%$ in model \eqref{whole model} under the null setting where $\theta^*$=0.}
	The external metabolite $\widetilde{m}_j$ was generated in the same way as the target metabolite.
	Under this data-generation regime, the 112 microbes were classified into the following four categories. 
	Specifically, 2 microbes belong to $\mathcal{G}_1$, 
	3 microbes belong to $\mathcal{G}_2$, 3 microbes in $\mathcal{G}_3$, and the remaining belong to $\mathcal{G}_4$. 
    While $m_i$ in the target data was generated to obtain the outcome $y_i$, we assume it is {\it not known} during estimation and inference.



\begin{figure}[H]
    \centering
    \includegraphics[width=0.6\linewidth]{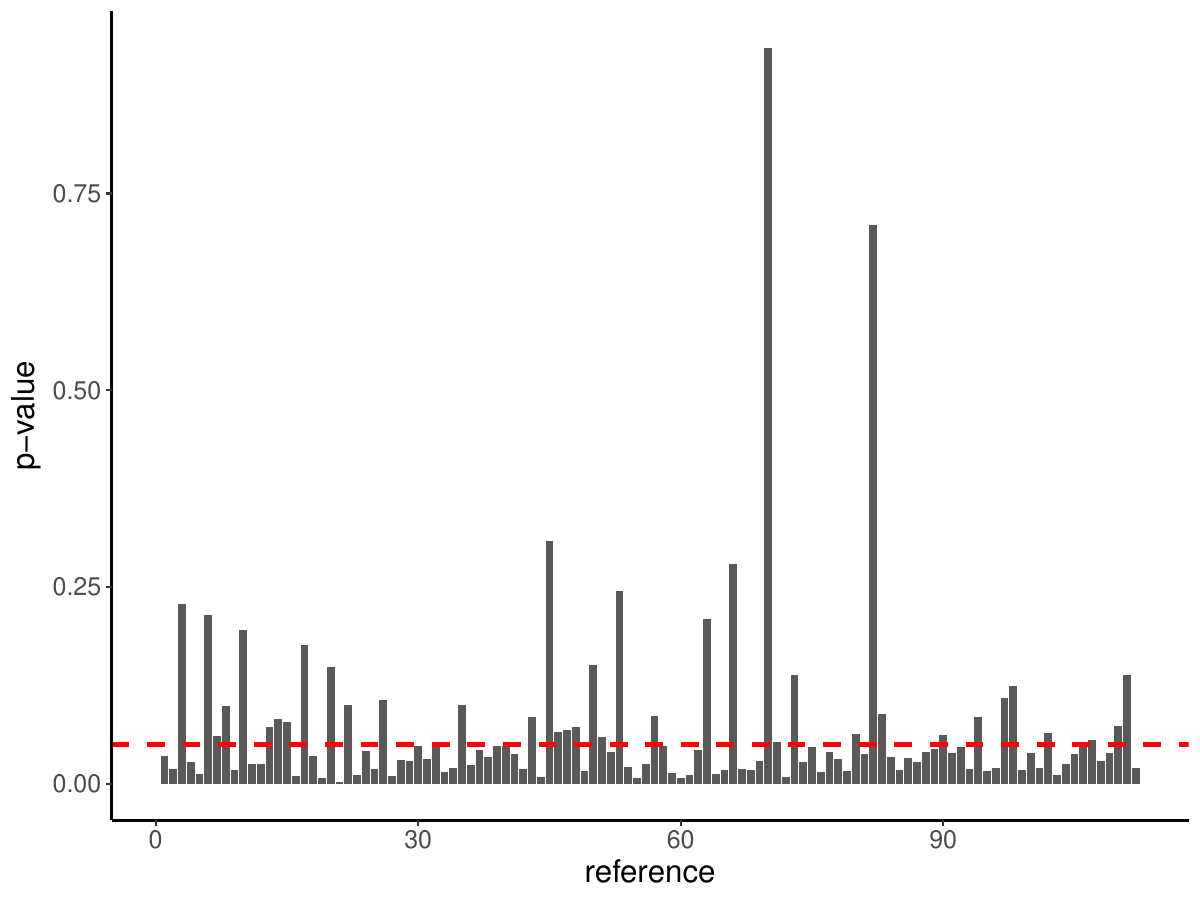}
    \small
    \caption[]
{{The distribution of p-values under different references when $N=220$, $n=347$, $\sigma_\delta=\sigma_\varepsilon=4$ and $\theta^*$=0.2. The horizontal red dashed line represents the p-value threshold of 0.05. There are 38 out of 112 references that yield p-values above 0.05.}}
    \label{fig: p-value-reference}
\end{figure}

{We first examined the robustness regarding the choice of the reference microbe in the ALR procedure.
Specifically, we set $N=220$, $n=347$, and $\theta^*=0.2$. For each possible reference taxon in the ALR transformation, we fitted the proposed model and obtained the p-value for testing $H_0:\theta^*=0$.
Fig. \ref{fig: p-value-reference} shows the distribution of p-values obtained using different reference taxa. The p-values range from 0.0024 to 0.9345, with a median of 0.0385. Moreover, approximately one-third of the references (38 out of 112) yield p-values greater than 0.05. These results clearly demonstrate that the choice of reference in the ALR transformation can substantially affect statistical inference.}

\begin{figure}
		\renewcommand{\thesubfigure}{\textsf{\Alph{subfigure}}}
		\centering
		\vspace{-5mm}
		\begin{subfigure}{0.48\textwidth}
			\captionsetup{justification=raggedright,singlelinecheck=false}
			\caption{}%
			\centering
			\includegraphics[width=\textwidth]{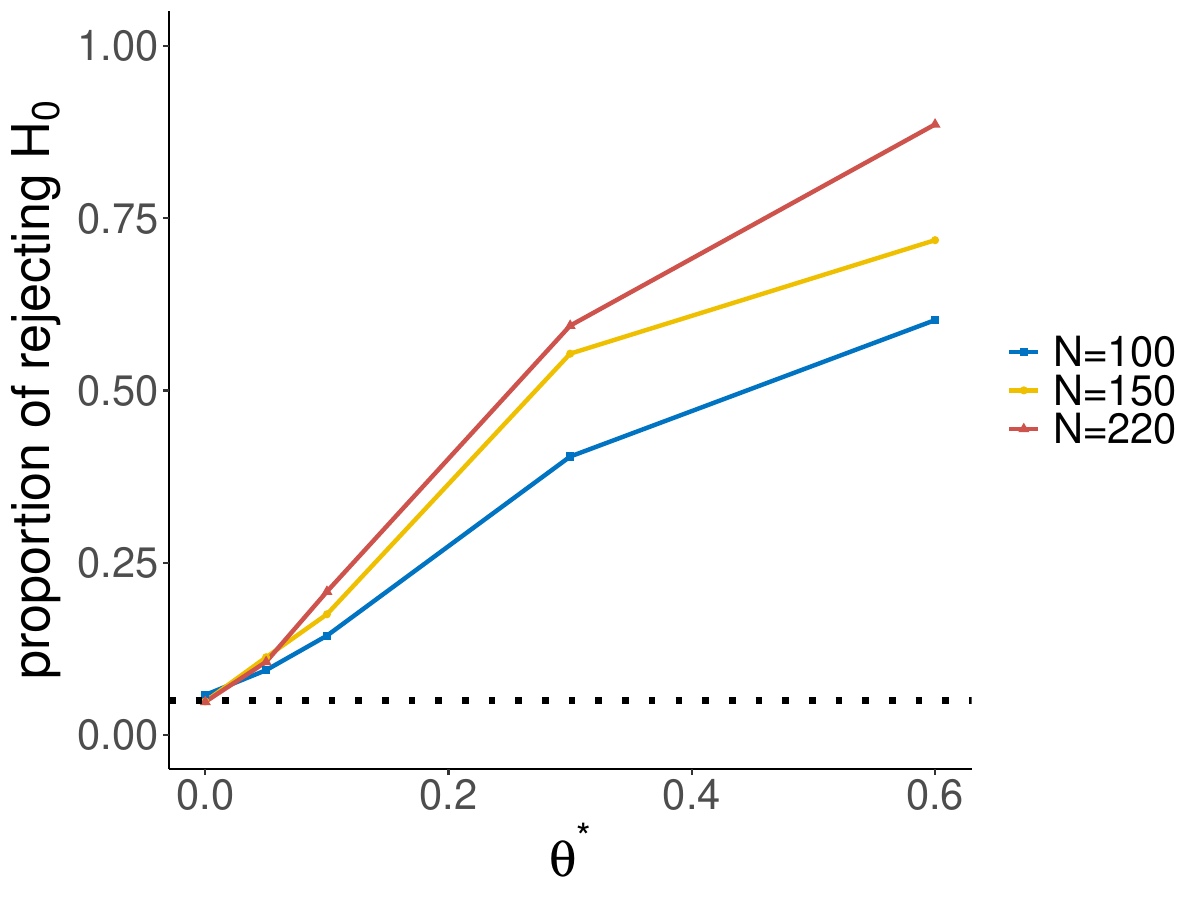}   
			\label{fig: lineplot-train-power}
		\end{subfigure}
		\begin{subfigure}{0.48\textwidth}  
			\captionsetup{justification=raggedright,singlelinecheck=false}
			\caption{}%
			\centering 
			\includegraphics[width=\textwidth]{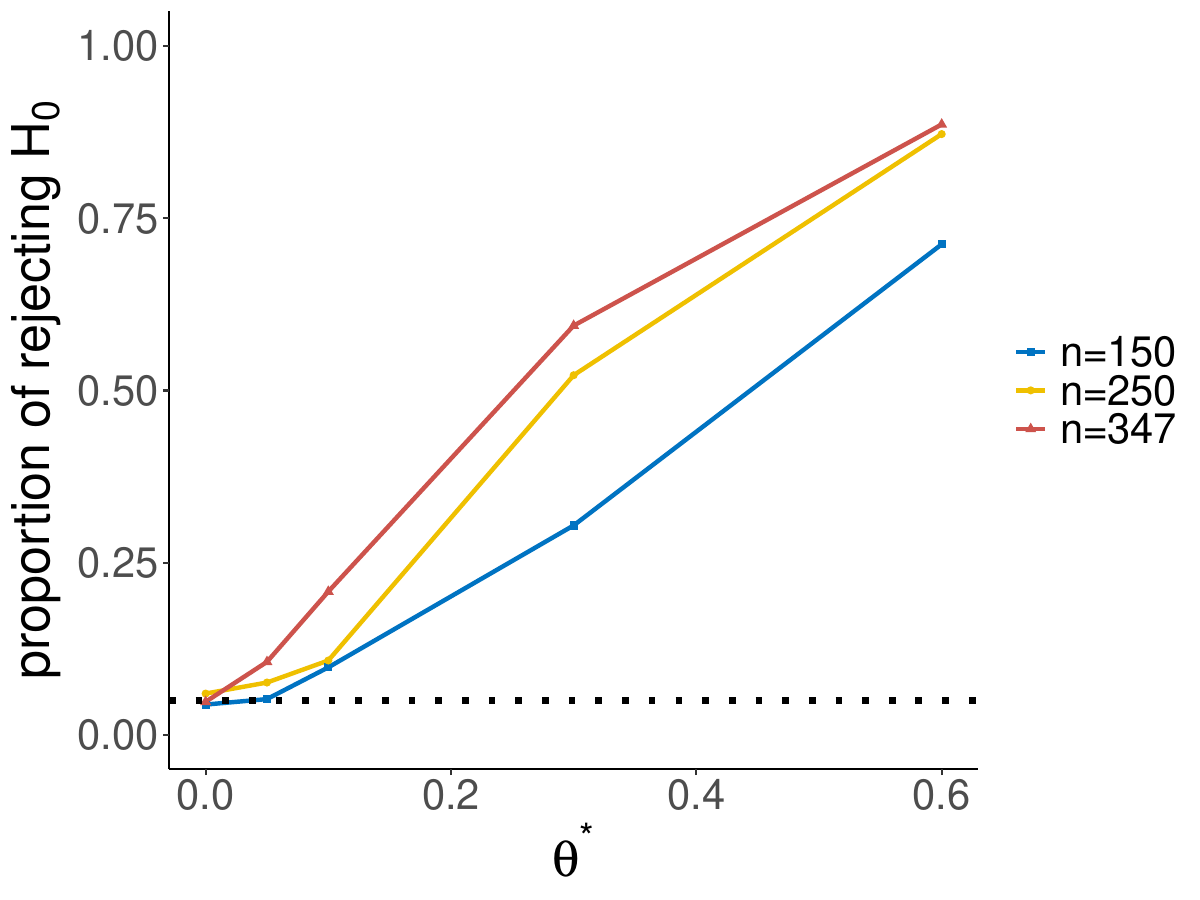}  
			\label{fig: lineplot-test-power}
		\end{subfigure}

        		\begin{subfigure}{0.48\textwidth}
			\captionsetup{justification=raggedright,singlelinecheck=false}
			\caption{}%
			\centering
			\includegraphics[width=\textwidth]{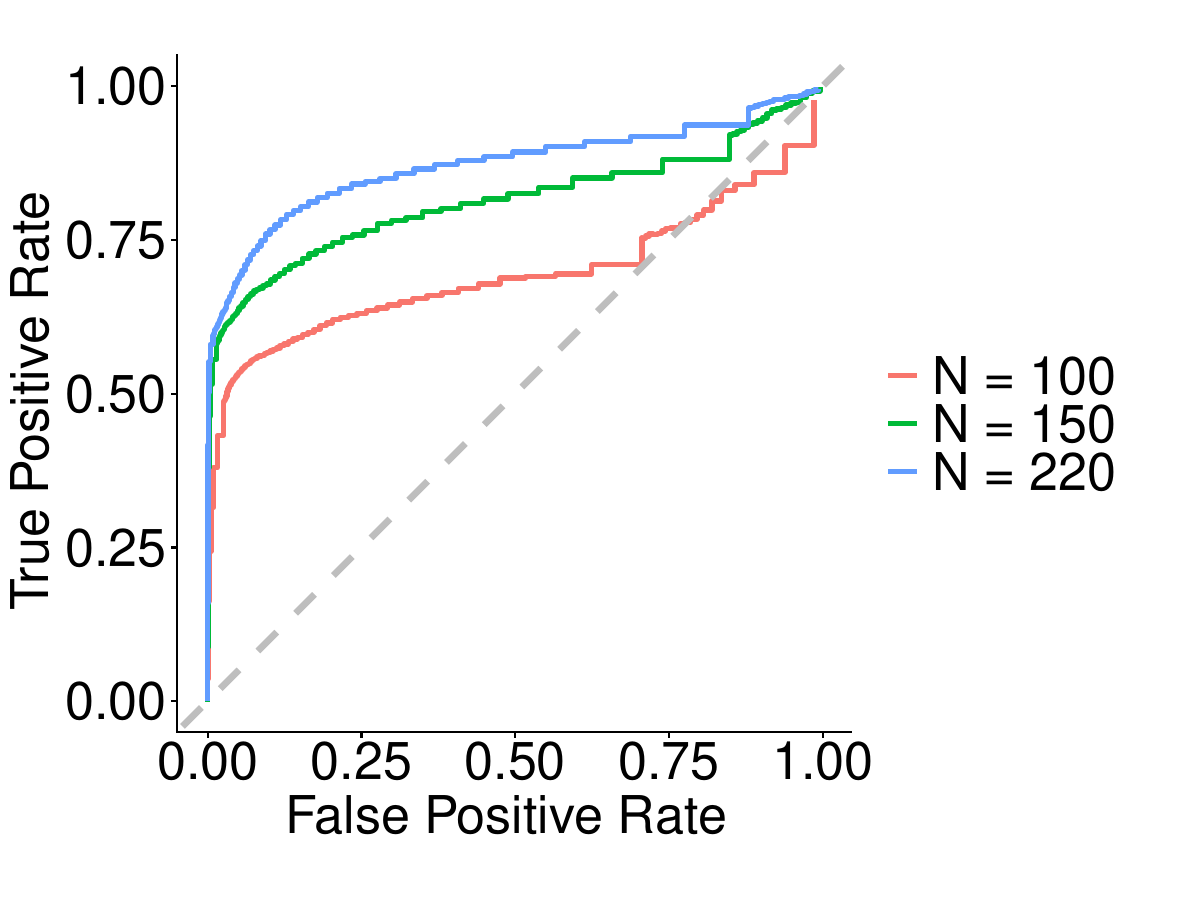}   
			\label{fig: ROC-train}
		\end{subfigure}
		\begin{subfigure}{0.48\textwidth}  
			\captionsetup{justification=raggedright,singlelinecheck=false}
			\caption{}%
			\centering 
			\includegraphics[width=\textwidth]{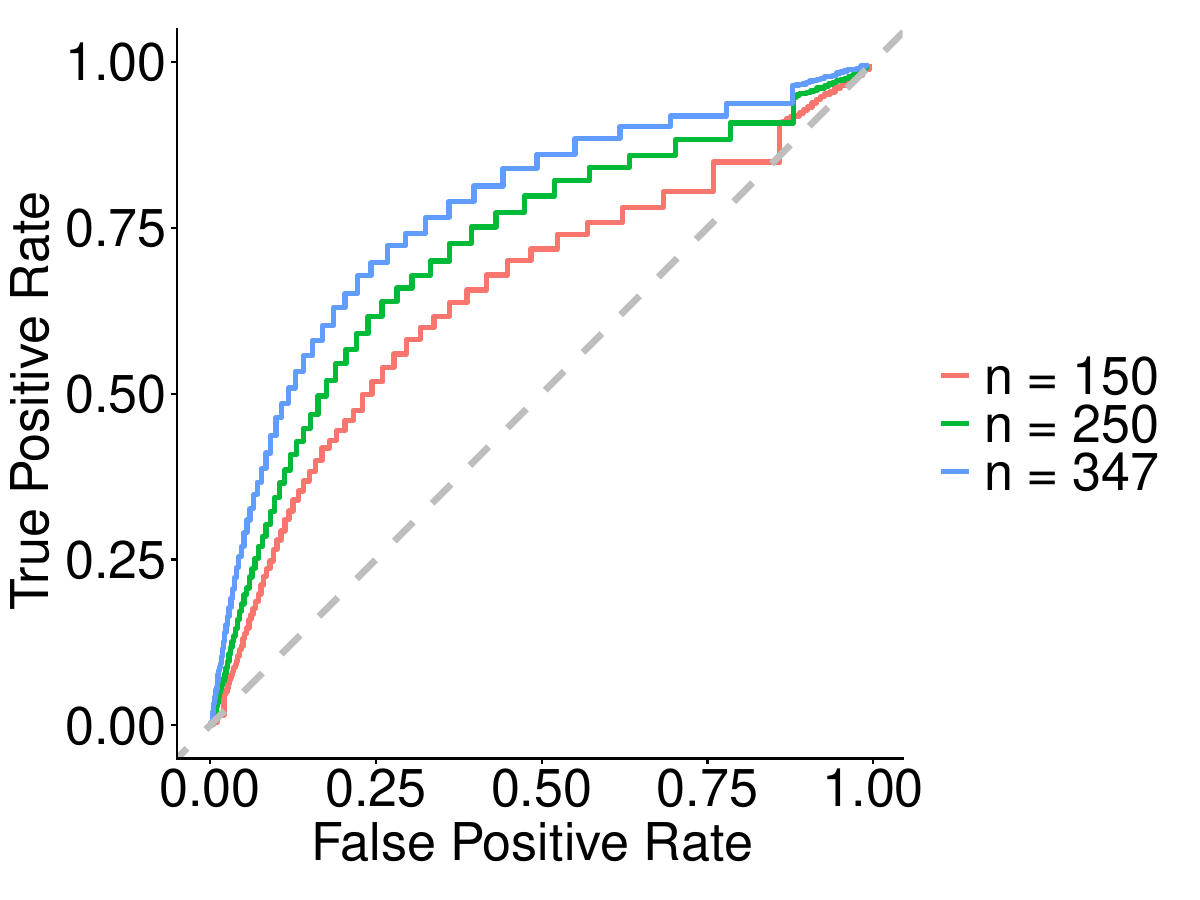}  
			\label{fig: ROC-test}
		\end{subfigure}  
        
		\caption[]
		{\small
    {Type I error and power for testing $H_0:\,\theta^* = 0$ when $\sigma_\delta=\sigma_\varepsilon=4$ under varying sample sizes. 
{(A)} External sample size varies; 
{(B)} Target sample size varies. 
The $R^2$ is approximately $23\%$ in stage 1 model \eqref{sub model}, and the $R^2$ is approximately $9\%$ in model \eqref{whole model} under the null setting where $\theta^*$=0.
The horizontal dotted line indicates the nominal level of 0.05.
{(C)} ROC curves for $\boldsymbol{\gamma}^0$ under different external sample sizes, with AUCs of 0.687, 0.815, and 0.870 for $N=100$, 150, and 220, respectively. 
{(D)} ROC curves for $\boldsymbol{\beta}^0$ under different target sample sizes, with AUCs of 0.656, 0.718, and 0.771 for $n=150$, 250, and 347, respectively.}
        }
		\label{fig: sd=4}
	\end{figure}

Next, we considered testing $H_0: \theta^* = 0$ using the proposed multi-reference procedure by 
setting $\theta^* = 0,0.05,0.1,0.3,{0.6}$. 
We used the full external sample size of $N=220$, while to examine the impact of the target sample size, we considered the full sample size of $n=347$ as well as two subsamples with $n=150$ and $250$. 
Additionally, to investigate the impact of the external sample size, we used the full target sample size of $n=347$ and set the external sample size of $N=100, 150, 220$.  The simulation results were examined based on 500 replications.
With varying external sample sizes, Fig. \ref{fig: lineplot-train-power} shows that the proposed method controls Type I error well and has increased power as external sample size increases. Fig.  \ref{fig: lineplot-test-power} shows a similar pattern as target sample size varies.
    
We then examined the categorization of microbial roles into the 4 groups by fixing $\theta^*=0.6$.  
Fixing the stability selection frequency ($\tau$) at $0.8$, we calculated the true positive rate (TPR) and false positive rate (FPR) according to 
\begin{align*}
\mbox{TPR}_{\lambda_\beta}(\mbeta^0) = \frac{ \sum_{j=1}^{p} \mathbbm{1}(\beta_j^0 \neq 0 \mbox{ and } \pi_j^{(\widehat{\mbeta})}>\tau) }{\sum_{j=1}^{p} \mathbbm{1}(\pi_j^{(\widehat{\mbeta})}>\tau) }, ~
\mbox{FPR}_{\lambda_\beta}(\mbeta^0) = \frac{ \sum_{j=1}^{p} \mathbbm{1}(\beta_j^0 = 0 \mbox{ and } \pi_j^{(\widehat{\mbeta})}>\tau) }{\sum_{j=1}^{p} \mathbbm{1}(\pi_j^{(\widehat{\mbeta})}>\tau) },
\end{align*}
where $p=112$, and $\lambda_\beta$ is the tuning parameter in \eqref{lasso: 2nd stage} used for estimating $\widehat{\bm{\beta}}$. 
We then showed the ROC curve by varying the tuning parameter $\lambda_\beta$. 
Subsequently, the ROC curve for $\mgamma^0$ were plotted in the same way.  
In Fig. \ref{fig: ROC-train}, all ROC curves lie above the diagonal reference line, indicating that our robust multi-reference procedure can accurately identify the non-zero entries of ${\bm{\beta}}^{0}$ across all external sample sizes.  
Moreover, the ROC curves move farther from the diagonal reference line as external sample size increases, indicating improved performance. 
This trend is also supported by the area under the curve (AUC), which increased from {0.687} to {0.815} and {0.870} for $N$ = 100, 150 and 220, respectively. 
Similar patterns are shown in Fig. \ref{fig: ROC-test} where the target sample size varies.
The AUC values in Fig. \ref{fig: ROC-test} show a improved performance with target sample size, increasing from {0.656} to {0.718} and {0.771} for $n$ = 150, 250 and 347, respectively.
Thus, Fig. \ref{fig: ROC-train} and \ref{fig: ROC-test} jointly demonstrate the effectiveness of the proposed multi-reference stability procedure for recovering $\mathcal{G}_j$ for $j=1,\ldots, 4$.
{
We also conducted a sensitivity analysis against the stability threshold. Following the original stability selection framework of \citep{meinshausen2010stability} and its implementation in the R package {\tt stabs}, the selection threshold is typically chosen within the range of 0.6--0.9. Hence, we additionally consider thresholds of 0.6, 0.7, and 0.9 in the simulation study while keeping all other simulation settings the same. We report the variable selection performance for recovering the nonzero entries of $\mgamma^0$ and $\mbeta^0$ under the different sample size settings.  Fig. S3 in Supplementary Materials Section 1.2 shows that the variable selection performance is robust to the choice of stability threshold, as evidenced by the nearly overlapping ROC curves across all thresholds considered. And this robustness persists under all sample size settings.  These findings are consistent with the general conclusions on the robustness of stability selection reported in \citep{meinshausen2010stability}}.

{Finally, to examine the robustness of proposed method under different signal-to-noise ratios, we considered additional case where $\sigma_\delta=\sigma_\varepsilon\in\{3,5\}$, while keeping the rest of the parameters the same. Accordingly, in model \eqref{sub model}, $R^2$ is approximately $34\%$ and $16\%$, respectively; 
in model \eqref{whole model}, under the null setting where $\theta^*$=0, $R^2$ is approximately $15\%$ and $6\%$, respectively. The simulation results are presented in Figs. S1 and S2 of Supplementary Materials Section 1.1. Consistent with Fig. \ref{fig: sd=4}, both figures show a similar patterns of Type I error, power and AUC for selecting $\mgamma^0$ and $\mbeta^0$ across different sample sizes.
In addition, Tables S1, S2, and S3 in the Supplementary Materials Section 1.1 summarize the Type I error, power (with $\theta^*=0.6$), and the AUC for selecting $\mgamma^0$ and $\mbeta^0$, respectively, under different values of $\sigma_\delta$ and $\sigma_\varepsilon$. Across all settings, the proposed test maintains satisfactory Type I error control, with empirical Type I error remaining close to the nominal 0.05 level. As expected, increasing the signal-to-noise ratio (equivalently, increasing $R^2$ by reducing $\sigma_\delta$ and $\sigma_\varepsilon$) substantially improves power. For example, when the target sample size increases from $N=100$ to $N=220$ with $n=347$ fixed, the power increases from 0.816 to 0.952 when $\sigma_\delta=\sigma_\varepsilon=3$, and from 0.474 to 0.788 when $\sigma_\delta=\sigma_\varepsilon=5$. Similarly, increasing the external sample size from $n=150$ to $n=250$ with $N=220$ fixed consistently improves power across all noise levels.}
{
The improvement in signal-to-noise ratio also enhances variable selection performance. As the noise level decreases, the AUCs for recovering both $\mgamma^0$ and $\mbeta^0$ increase consistently. Moreover, larger target and external sample sizes further improve selection accuracy. For example, the AUC for recovering $\mgamma^0$ increases from 0.732 to 0.921 as $N$ increases from 100 to 220 when $\sigma_\delta=\sigma_\varepsilon=3$, while the AUC for recovering $\mbeta^0$ increases from 0.736 to 0.843 as $n$ increases from 150 to 347 under the same noise level. Notably, even under the highest noise level considered ($\sigma_\delta=\sigma_\varepsilon=5$), the proposed method still achieves reasonably good signal recovery, with AUCs above 0.60.}

\subsection*{Real Data Application}\label{sec: real data}
	In this section, we aim to elucidate microbiome-metabolome {relationships} related to the disease pathology of IBD by 
	integrating iHMP (external) \citep{lloyd2019multi} and the FRAN (target) dataset. 
	The FRAN cohort consists of 220 individuals (164 IBD cases), while the iHMP cohort consists of 105 individuals (79 cases). Since iHMP is a longitudinal study, we only use the baseline measurement defined by the first visit for each individual. The metagenomics and metabolomics data in both datasets were collected from stool samples by shotgun and four LC-MS methods, respectively.

	\subsubsection*{Data filtering} \label{sec:rd:pre}
	The iHMP and FRAN datasets have been preprocessed and curated using the same pipelines \citep{muller2022gut}. 
	For the FRAN study, the curated microbiome data contain 9,974 bacterial genera. The curated metabolomic data include 8,849 metabolites, of which 149 metabolites have annotated HMDB IDs with a high confidence level. The curated iHMP dataset contains 7,235 bacterial genera and 81,868 metabolites, where 272 metabolites have annotated HMDB IDs with a high confidence level. 
	
	We first processed the microbiome abundance in both datasets. 
    Fig. \ref{fig: pre-process} summarizes the main steps. Specifically, we aggregated the FRAN microbiome abundance from bacterial genera into 1,140 bacterial orders, aiming to reduce dimensionality. 
		Similarly, for the iHMP microbiome abundance, after aggregation, we obtained 835 bacterial orders, 798 of which are also measured in FRAN dataset. 
        We adopted R package {\tt MMUPHin} \citep{ma2022population} to perform batch-effect correction on the 798 overlapping bacterial orders, adjusting for study group (case vs control) and age.
        One subject from the iHMP dataset was excluded due to missing age information.
        After batch effect correction, we have two steps of handling zeros in the microbiome data. First, we removed microbial orders with more than 30\% of zeros across participants in FRAN dataset. 
        Among them, 86 microbial orders overlapped with the iHMP dataset and were taking into the remaining analysis.  {Second, we added a pseudo-abundance to the remaining taxa before fitting the proposed model.
A common approach for handling zeros in log-contrast models is to add a small pseudo-abundance (e.g., 1 or 0.5) when absolute counts are available \citep{aitchison1986}. However, only relative abundance data were available in our analysis, and the corresponding raw counts were unavailable. 
Therefore, we added a pseudo abundance of $10^{-8}$, which is smaller than the minimum observed nonzero relative abundance ($7.28\times10^{-8}$), to minimize its impact on the observed measurements in both FRAN  and iHMP datasets for the ALR transformation.}

	For the 272 annotated metabolites in the iHMP dataset, we removed those with any missing values, resulting in 143 metabolites left. 
    Among these, 113 metabolites were completely measured only in the iHMP dataset, while only 30 were shared between the two datasets. 
    This highlights the need to integrate FRAN and iHMP for a more comprehensive metabolomic analysis.
  Because our primary interest is in microbiome–metabolite–IBD pathways, we restricted our analysis to 76 metabolites that were related to the microbiome. 
  These microbially associated metabolites were identified using {\tt MiRKAT} \citep{zhao2015testing}, with Bray--Curtis dissimilarity \citep{bray1957ordination} applied to the microbiome data and log-transformed metabolite abundances. 
  Specifically, 59 of the 113 iHMP-unique metabolites and 17 of the 30 metabolites shared across datasets are identified as significant. 
  

We selected FRAN as our target dataset because it has a larger sample size than iHMP and includes fecal calprotectin (FCP) measurements, a clinically relevant IBD outcome \citep{moniuszko2013biomarkers}. 
Individuals in FRAN with missing FCP measurements were excluded.
After filtering, the final target dataset (FRAN) comprised 153 subjects (111 cases) and 86 microbial orders. The external dataset (iHMP) included 104 subjects (78 cases) with the same 86 microbial orders and 76 microbially related metabolites, 17 of which were shared with FRAN.



		\subsubsection*{Transferability assessment} 
    
\begin{figure}
		\renewcommand{\thesubfigure}{\textsf{\Alph{subfigure}}}
		\centering
      \begin{subfigure}{\textwidth}
			\captionsetup{justification=raggedright,singlelinecheck=false}
            	\caption{}%
			\centering
     \includegraphics[width=1\textwidth]{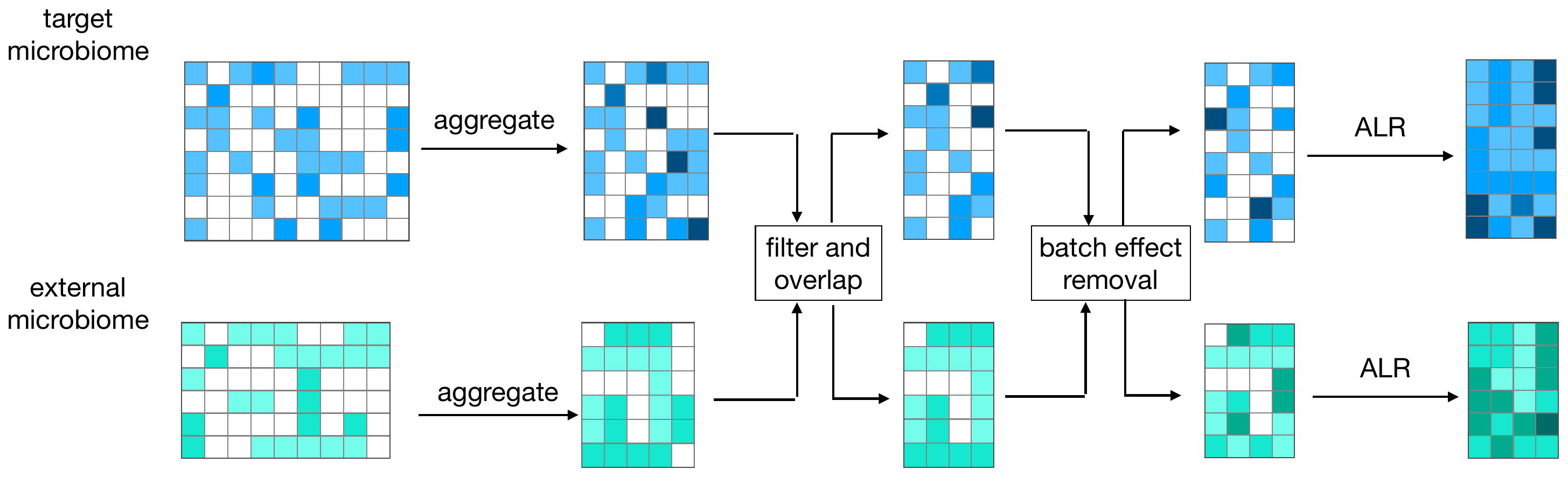} 
		\label{fig: pre-process} 
		\end{subfigure}
              \begin{subfigure}{0.31\textwidth}
			\captionsetup{justification=raggedright,singlelinecheck=false}
            	\caption{}%
			\centering
     \includegraphics[width=\textwidth]{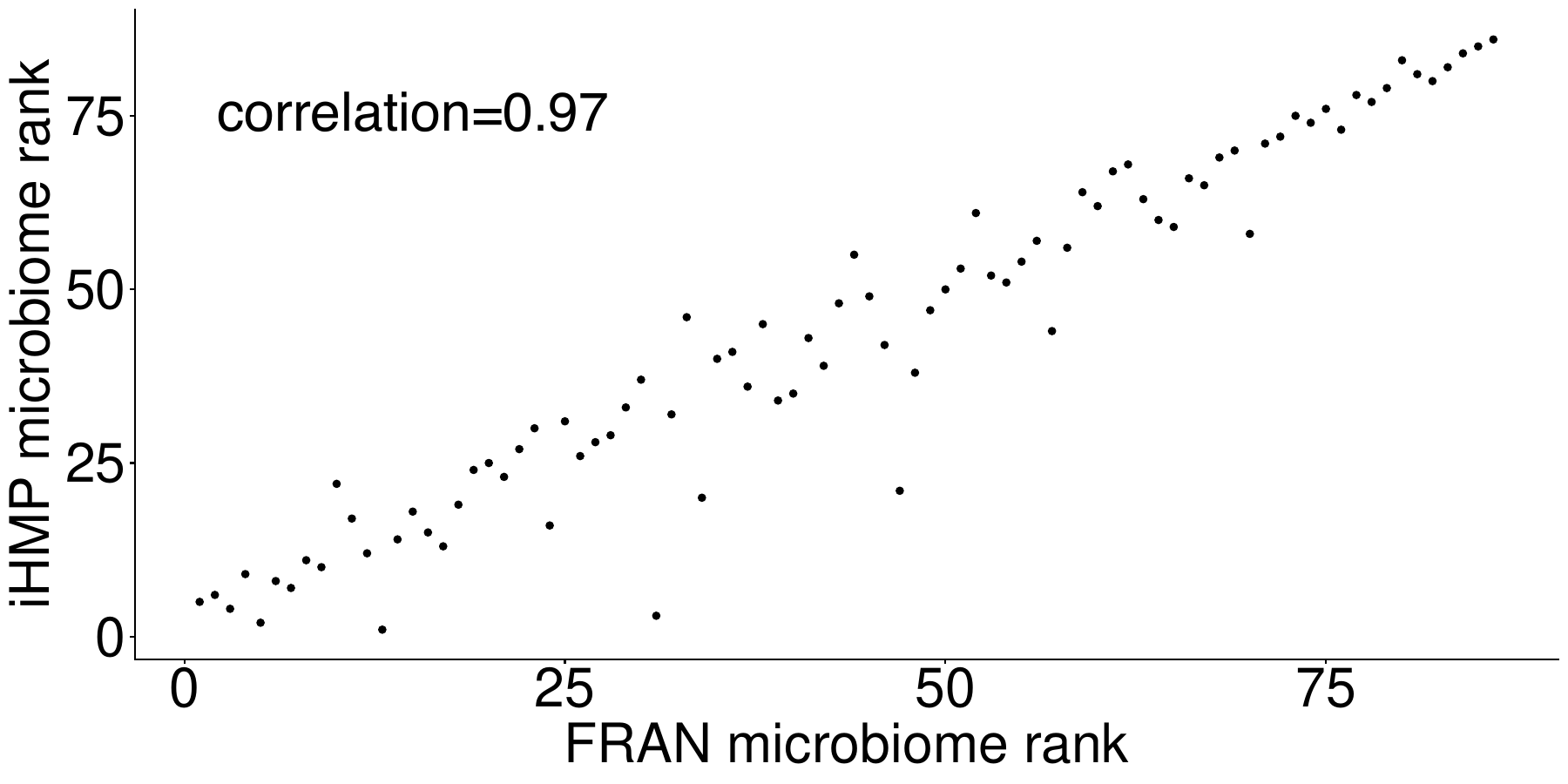} 
		\label{fig: rank-correlation} 
		\end{subfigure}
        		\begin{subfigure}{0.31\textwidth}
			\captionsetup{justification=raggedright,singlelinecheck=false}
			\caption{}%
			\centering
			\includegraphics[width=\textwidth]{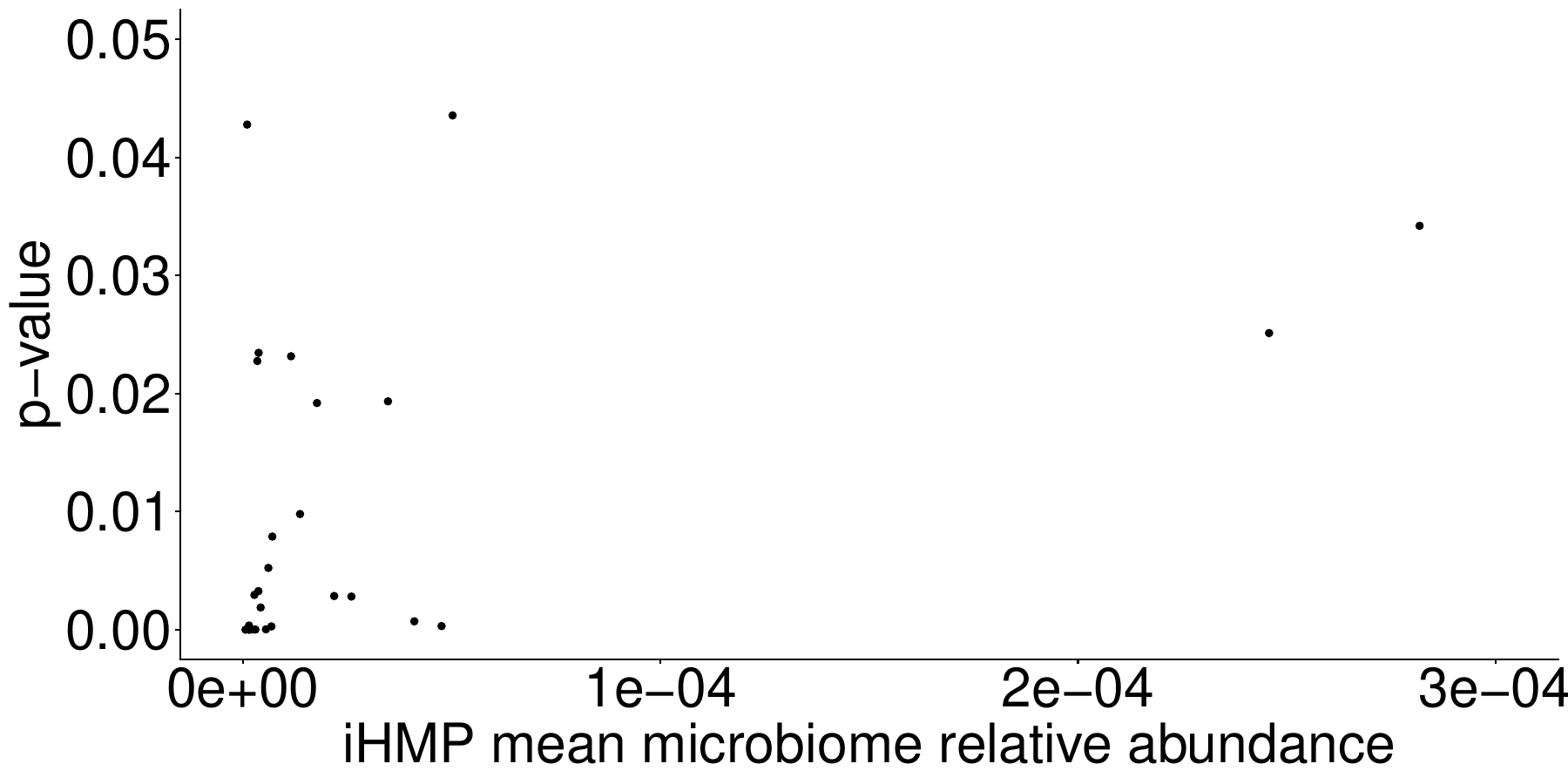}   
			\label{fig: iHMP-mean-pvalue}
		\end{subfigure}
		\begin{subfigure}{0.31\textwidth}
			\captionsetup{justification=raggedright,singlelinecheck=false}
			\caption{}%
			\centering
			\includegraphics[width=\textwidth]{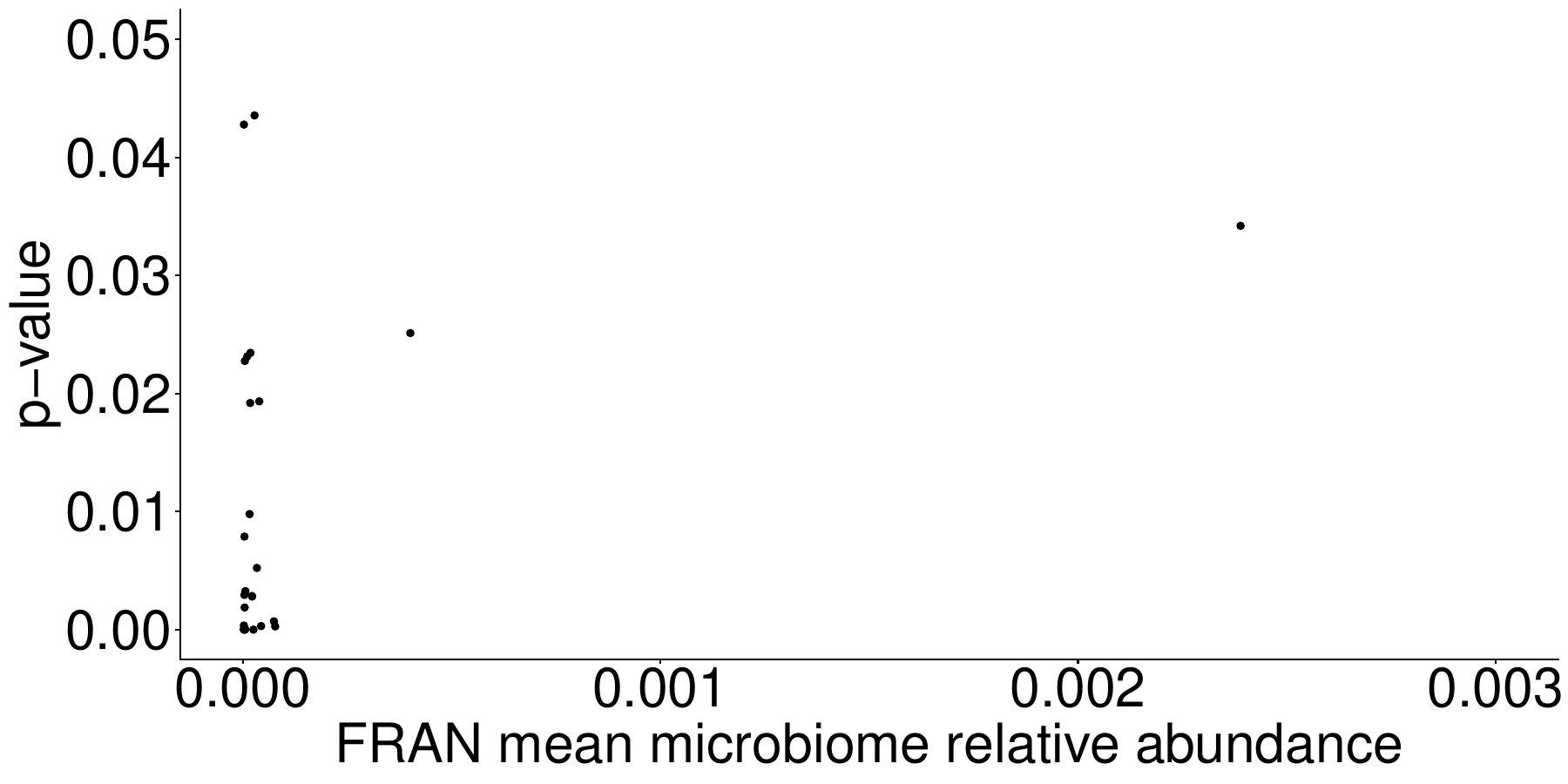}   
			\label{fig: FRAN-mean-pvalue}
		\end{subfigure}

		\begin{subfigure}{0.48\textwidth}
			\captionsetup{justification=raggedright,singlelinecheck=false}
			\caption{}%
			\centering
			\includegraphics[width=\textwidth]{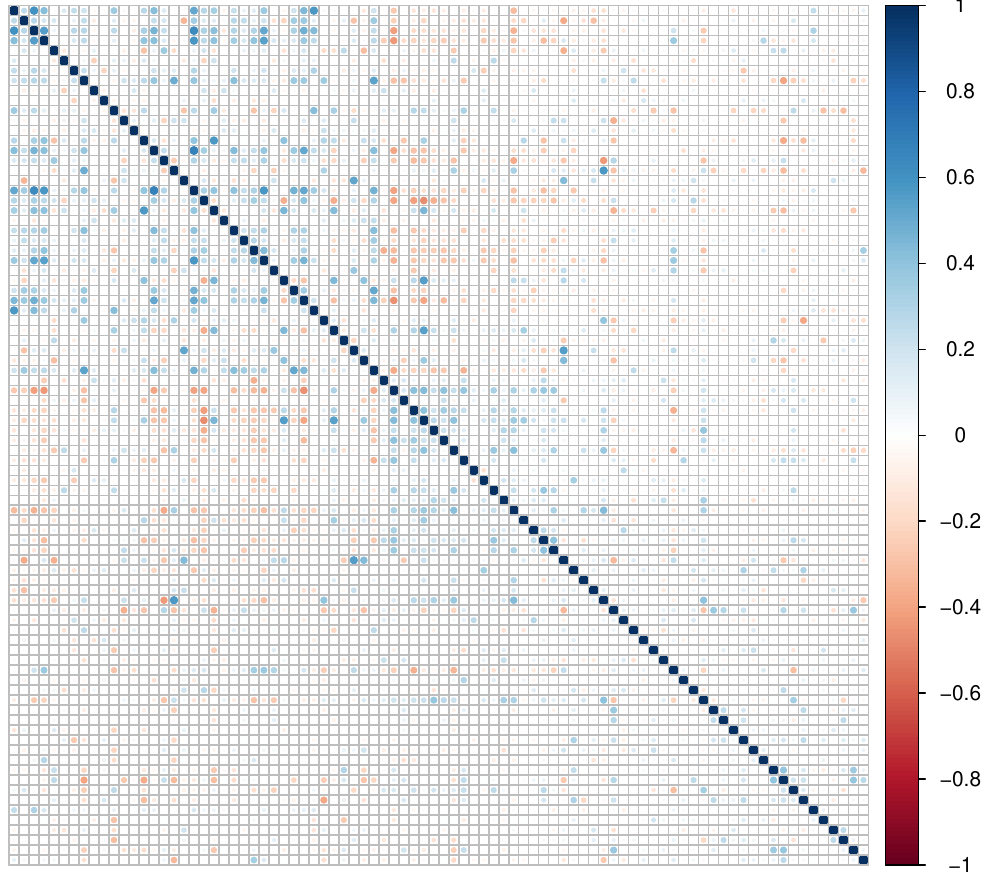}   
			\label{fig: cor-micro-IHMP}
		\end{subfigure}
		\begin{subfigure}{0.48\textwidth}  
			\captionsetup{justification=raggedright,singlelinecheck=false}
			\caption{}%
			\centering 
			\includegraphics[width=\textwidth]{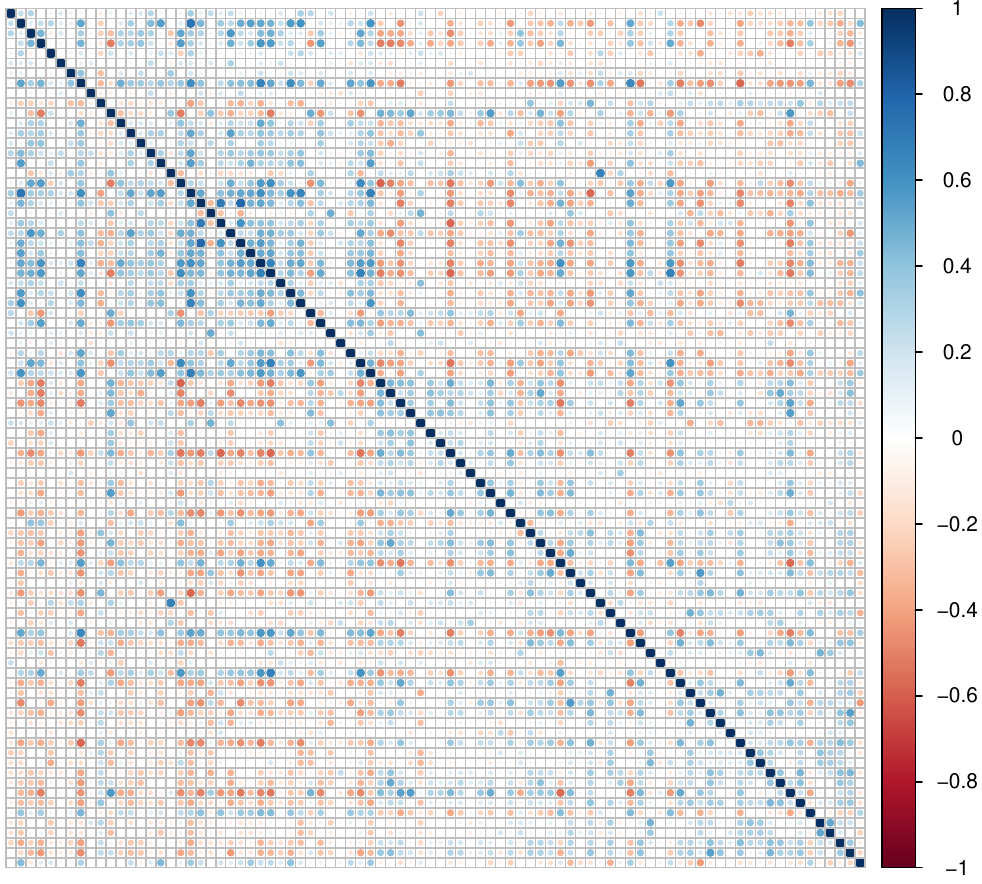}  
			\label{fig: cor-micro-FRAN}
		\end{subfigure}

		\caption[]
		{\small 
       (A) Processing pipeline for microbiome relative abundance data in the iHMP and FRAN studies.
(B) Scatter plot showing highly concordant microbiome abundance ranks between the FRAN and iHMP datasets.
(C) and (D) Scatter plots illustrating that microbes with discordant ranks between FRAN and iHMP tend to be low-abundance taxa in the iHMP and FRAN datasets, respectively.
Correlation matrix plots of microbiome relative abundance in the iHMP (E) and FRAN (F) datasets, showing similar structures.
        }
		\label{fig: transferbility-assessment}
	\end{figure}

	Our framework assumes that information from the external dataset (iHMP) can be reliably transferred to the target dataset (FRAN). 
    To support this assumption, we conducted several descriptive analyses to assess cross-dataset similarity. 
    First, we compared the overall microbiome composition between iHMP and FRAN by computing, for each microbe, the average relative abundance across subjects in each dataset. 
    We then ranked these average relative abundances and examined their relationship, as shown in Fig.~\ref{fig: rank-correlation}. 
    The strong concordance between the two rankings, with a correlation of 0.97, indicates that the microbial communities exhibit highly similar overall abundance patterns across the two datasets.

Recognizing the heterogeneity across subjects, we extended this analysis to the individual level. 
Specifically, within each individual, we ranked the relative abundances of microbes, yielding two rank matrices corresponding to the two datasets. 
We then compared the rank of each microbe between the two datasets using the Wilcoxon rank-sum test. 
Among the 86 microbes examined, 28 showed marginal significance, and only 10 remained significant after Bonferroni correction. 
Notably, all 28 marginally significant microbes were among the lowest-abundance taxa: 
their largest average relative abundances were $2.82 \times 10^{-4}$ in the iHMP dataset and $2.39 \times 10^{-3}$ in the FRAN dataset. 
Fig.~\ref{fig: iHMP-mean-pvalue} and ~\ref{fig: FRAN-mean-pvalue} further illustrate the relationship between marginal p-values and average relative abundance of those 28 significant taxa, 
showing that the most significant microbes tended to have extremely low abundance, 
which is known to be more variable across subjects and reflects underlying biological heterogeneity.

        
	Since our SEM is a multivariable regression model that simultaneously includes all taxa as covariates, 
the correlation structure of microbiome abundances is crucial for reliable data transfer. 
We therefore used CCLasso \citep{fang2015cclasso}, a compositionality-aware correlation estimation method, 
to compare the taxon–taxon correlation matrices in the iHMP and FRAN datasets. 
As shown in Figs.~\ref{fig: cor-micro-IHMP} and~\ref{fig: cor-micro-FRAN}, 
the two correlation matrices exhibit similar overall block structures, 
with the FRAN dataset showing slightly stronger correlations among taxa. 
We further applied the Mantel test to confirm the similarity between the two matrices, 
which yielded a p-value $=0.001$.

Finally, we compared the log-transformed abundances of the 17 shared metabolites between the two datasets using two-sample $t$-tests. 
Based on marginal p-values, we found no significant differences in metabolite abundances across the two datasets. 
This result suggests substantial similarity between the shared metabolomic profiles of the two datasets.

	\subsubsection*{Integration of FRAN and iHMP data}\label{sec: rd: int}
In this section, we applied the proposed multi-reference procedure (Algorithm 1) to integrate the target FRNA dataset ($n=153$) with the external iHMP dataset ($n=104$). After data filtering, 86 microbial orders and 76 microbiome-related metabolites were retained. 
Our goal was to identify biologically meaningful microbiome–metabolome pathways associated with IBD. 
Metabolite levels were standardized to have mean zero and unit variance in the external dataset, while the FCP outcome was centered in the target dataset. In both datasets, the microbiome abundance matrices were centered after applying the ALR transformation.   
When implementing the penalized two-stage squared algorithm, we hard-thresholded each entry of $\widehat{\bm{\gamma}}$ in \eqref{lasso: 1st stage} at $10^{-10}$ to avoid numerical instability caused by extremely small Lasso estimates.

\begin{figure}
		\renewcommand{\thesubfigure}{\textsf{\Alph{subfigure}}}
		\centering
      \begin{subfigure}{\textwidth}		\captionsetup{justification=raggedright,singlelinecheck=false}
            	\caption{}%
\includegraphics[width=0.8\textwidth]{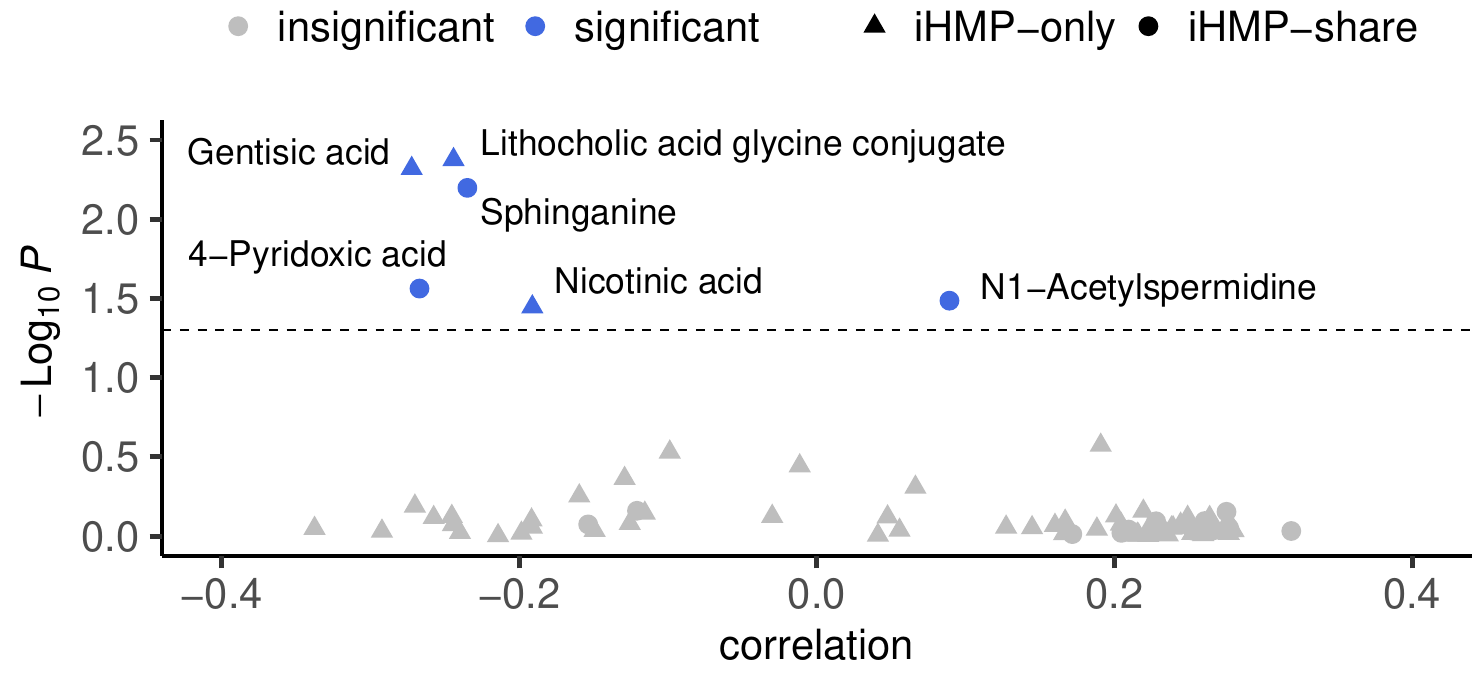} 
		\label{fig: volcano} 
		\end{subfigure}

		\begin{subfigure}{\textwidth}
			\captionsetup{justification=raggedright,singlelinecheck=false}
			\caption{}%
			\centering
			\includegraphics[width=0.7\textwidth]{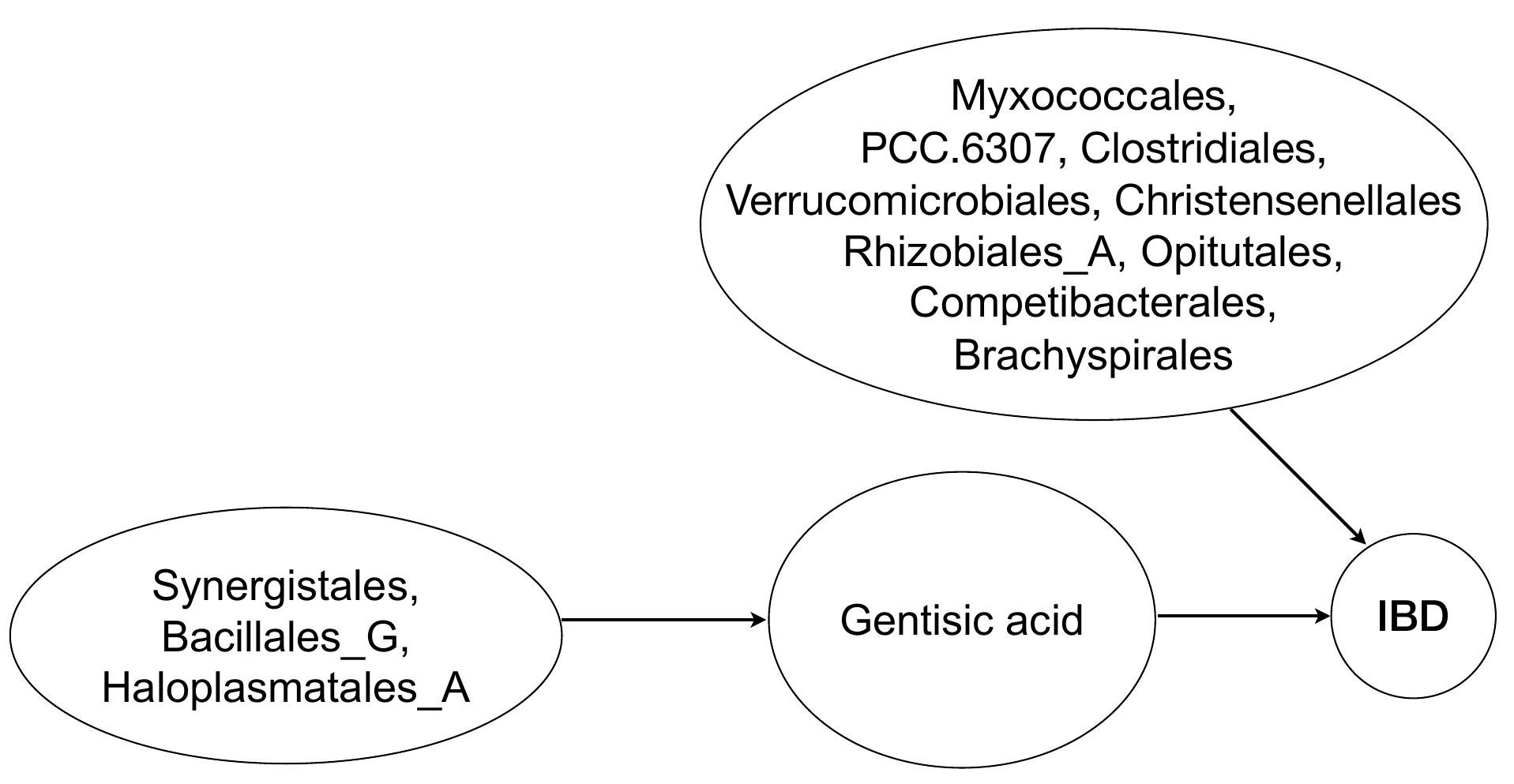} 
		\label{fig: diagram} 
		\end{subfigure}

		\caption{
(A) Volcano plot showing three IBD-associated metabolites identified by our multi-reference procedure. 
Triangle dots represent metabolites measured only in the iHMP dataset (iHMP-only), while solid circles represent metabolites shared between the iHMP and FRAN datasets (iHMP-share). 
Blue dots indicate significant metabolites ($p < 0.05$), and grey dots indicate non-significant metabolites. 
The dashed line marks the nominal significance level of $-\log_{10}(0.05)$.  
(B) Diagram of the microbiome–metabolome–IBD pathway for gentisic acid. 
Twelve microbial taxa were identified as participating in this pathway and are grouped into two functional categories.
}

		\label{fig: IBD pathway}
	\end{figure}

We first present a volcano plot summarizing the Cauchy-combined p-values ($\widetilde{p}$) and the median correlation statistics ($\widetilde{T}$) for all 76 metabolites (Fig.~\ref{fig: volcano}). 
Three metabolites showed marginal significance: two were among the 17 metabolites shared between FRAN and iHMP, and one was unique to iHMP. 
None remained significant after controlling the false discovery rate at 0.05, likely due to limited signal strength.
We then conducted a secondary analysis restricted to the 17 shared metabolites using only the target dataset (FRAN), fitting \eqref{ALR whole model} directly with the observed metabolite abundances. 
Neither of the two previously identified shared metabolites reached marginal significance in this analysis.  
Although methodologically different, the contrast between these results suggests that our integrative framework may improve power for detecting microbiome--metabolome--IBD pathways. 
Additional details of the secondary analysis are provided in the Supplementary Materials.

The identified metabolites have been investigated in prior studies as candidate biomarkers for IBD. 
Sphinganine is a central precursor in sphingolipid biosynthesis. Sphingolipids regulate inflammatory signaling and immune cell functions, and their dysregulation has been directly implicated in pro-inflammatory pathways relevant to IBD pathogenesis \citep{sukocheva2020sphingolipids}.
N1-acetylspermidine is an acetylated polyamine associated with polyamine turnover. The polyamines are involved in both the differentiation of immune cells and in the regulation of the inflammatory reaction \citep{moinard2005polyamines}.
Gentisic acid can suppress inflammatory responses by controlling the production of nitric oxide and pro-inflammatory cytokines and modulating inflammation-related protein pathways \citep{kang2021modulation}.
{Lithocholic acid glycine conjugate is the glycine-conjugated form of lithocholic acid (LCA), a secondary bile acid produced by gut bacteria. Patients with IBD consistently exhibit reduced levels of LCA, reflecting impaired microbial conversion of primary bile acids \citep{thomas2022emerging}. 
Nicotinic acid activates GPR109A, which is an anti-inflammatory receptor \citep{singh2014activation}.
4-Pyridoxic acid is the major degradation product of vitamin B6 \citep{vrolijk2017vitamin} while significantly lower vitamin B6 levels was found in IBD patients than in healthy controls, suggesting disturbed vitamin B6 metabolism in IBD \citep{saibeni2003low}.
}
    
As an illustrative example, we present the microbiome--gentisic acid--IBD pathway in Fig.~\ref{fig: diagram}, with the remaining pathways reported in the Supplementary Materials.
We identify three microbial taxa that indirectly influence FCP through gentisic acid, suggesting a potential mediating role of this metabolite in the association between these microbes and IBD.
In contrast, nine microbial taxa exhibit direct effects on FCP that are independent of gentisic acid.
Together, this pathway illustration provides an integrative view of how gentisic acid--related microbial taxa may contribute to IBD pathology.

{
To assess the sensitivity of our results to the choice of pseudo-abundance, we conducted additional analyses using two alternative values: one-half of the minimum observed nonzero relative abundance ($3.64\times10^{-8}$) and $10^{-9}$.}
{
We then applied the proposed framework separately to each transformed dataset obtained by adding one of the three pseudo-abundances. For each analysis, we obtained a p-value for the 76 metabolite of interest. We then compared the resulting significance levels by constructing pairwise scatter plots of the corresponding $-\log_{10}(\textit{p})$ values and calculating the Pearson correlation coefficient for each pair, as shown in Fig. S10 in Supplementary Materials Section 2.3.}

{
Fig. S10 shows strong agreement among the three approaches for handling zeros. The highest correlation is observed between the results obtained using one-half of the minimum observed nonzero relative abundance and $10^{-8}$ (Pearson correlation = 0.91), followed by the comparison between $10^{-8}$ and $10^{-9}$ (0.86). The correlation between one-half of the minimum observed nonzero relative abundance and $10^{-9}$ is slightly lower (0.77), but still indicates substantial agreement. This pattern is consistent with the relative magnitudes of the pseudo-abundances: the two more similar pseudo-abundances yield the most similar inference results.} 

{
We further compared the metabolites identified as significant using a nominal significance level of 0.05. The overlap of significant metabolites across the three pseudo-abundance choices is shown in the Venn diagram in Fig. S11 in Supplementary Materials Section 2.3. Using one-half of the minimum observed nonzero relative abundance ($3.64\times10^{-8}$), $10^{-8}$, and $10^{-9}$ yielded 3, 6, and 6 significant metabolites, respectively. Among these, two metabolites were consistently identified by all three analyses.
Although the overall p-values are strongly correlated across different pseudo-abundance choices (Fig. S10), the sets of metabolites declared significant are less consistent. This discrepancy likely reflects the use of a fixed significance threshold, where relatively small changes in p-values for metabolites near the 0.05 cutoff can lead to different significance decisions. Nevertheless, these findings suggest that the final list of significant metabolites may be sensitive to the choice of zero-handling strategy. Developing inference procedures that are more robust to zero handling is therefore an important direction for future research.
}


\section*{Discussion}

In this paper, we propose a structural equation model (SEM) to integrate two complementary datasets and identify biologically meaningful microbiome–metabolome–host pathways. By design, our SEM is robust to unmeasured confounding, which often arises in microbiome–metabolite–host studies. While our approach is related to \citep{yu2022mapping}, their work focuses on jointly analyzing genetic–imaging–clinical pathways in a single complete dataset, addressing a different question and requiring distinct analytical tools.

A key innovation of our framework is the robust multi-reference procedure for log-contrast models. This eliminates the dependence on a specific reference in ALR-based regression with compositional microbiome data, a limitation commonly encountered in existing methods. 
{Recently, a new reference-free method is proposed for modeling the association between compositional data and an outcome \citep{li2023s}. However, to the best of our knowledge, it is not immediately clear how this methodology can be adapted to the high-dimensional two-stage least squares framework considered in our paper. Moreover, our primary focus is valid statistical inference for the mediation effect, whereas the applicability of \citep{li2023s} to this inference problem has not yet been established. Extending their methodology to our setting and investigating its inferential properties would be an interesting direction for future research.}

There are several avenues for future improvement. First, our regularized permutation test may be conservative because the penalization in Step 2 reduces the effect of predicted metabolites, potentially lowering statistical power. {An attractive alternative would be to apply an existing debiased-lasso method for high-dimensional inference \citep{van2014asymptotically,zhang2014confidence}. 
However, to the best of our knowledge, no existing debiasing procedure is applicable to the two-stage least squares framework with high-dimensional compositional covariates considered in this paper. Developing a debiasing procedure that is valid for our SEM framework would be an interesting direction for future research.}
Second, the Cauchy combination used to aggregate inference results across all references, while intended to increase power, can be overly influenced by a single extremely small p-value. Alternative inferential procedures may improve robustness and power. {Third, although a continuous outcome is considered for the present study, extending the proposed framework to binary outcomes would broaden its applicability. Our framework consists of two components: estimation and inference. The
estimation procedure can be naturally extended to binary outcomes by replacing the penalized least squares regression with penalized logistic regression. However, the proposed inference procedure was specifically developed for continuous outcomes. In particular, the permutation test relies on residuals to remove the effects of nuisance parameters while preserving the null hypothesis \citep{hemerik2021permutation}. Such residual-based permutation procedures are not directly applicable to binary outcomes because the response is discrete and the residuals no longer possess the same properties. Developing a valid and computationally efficient inference procedure for binary outcomes would therefore require additional methodological work and is an important direction for future research.}

Finally, when multiple external cohorts are available, efficiency can be improved by leveraging all of them rather than using a single dataset. One possible strategy is to weight external datasets according to their informativeness. For example, for metabolites observed in both the target and external datasets, we can compute the correlation between predicted and observed values in the target. The average correlation across metabolites can then serve as a weight for each external dataset, allowing a proportional integration of information from all sources.

	
	\section*{Supporting information}
	 Supplementary file with additional real data results, mathematical discussion of model assumptions, theoretical results, and proofs.
	
	\section*{Acknowledgments}
	Y.W is funded by  National Institutes of Health (NIH) R03OD039969. 


	\section*{Author Contributions}
    
	\textbf{Conceptualization:} Yue Wang.
    
    \noindent\textbf{Data curation:} Lei Fang.

    \noindent\textbf{Formal analysis:} Lei Fang.

\noindent\textbf{Funding acquisition:} Yue Wang.

\noindent\textbf{Investigation:} Lei Fang, Chenglong Ye, Yue Wang.

\noindent\textbf{Methodology:} Lei Fang, Yue Wang.

\noindent\textbf{Project administration:} Yue Wang.

\noindent\textbf{Resources:} Yue Wang.

\noindent\textbf{Supervision:}  Chenglong Ye, Yue Wang.

\noindent\textbf{Writing – original draft:} Lei Fang, Chenglong Ye.

\noindent\textbf{Writing – review \& editing:} Yue Wang.

	
	\nolinenumbers
	
	%
	%
	
	\bibliography{reference}

\end{document}